\documentclass[prd,twocolumn,floatfix]{revtex4}

\usepackage{amssymb}
\usepackage[dvips]{graphicx}
\usepackage{psfrag}
\usepackage{subfigure}

\begin{document}

\title{Static potential and local color fields in unquenched
lattice QCD$_3$}

\author{Howard D. Trottier and Kit Yan Wong}
\affiliation{Simon Fraser University, Department of Physics, 
8888 University Drive, Burnaby BC V5A 1S6, Canada}

\begin{abstract}
String breaking by dynamical quarks in three-dimensional lattice QCD is 
analyzed through measurements of the potential and the local 
color-electric field strength generated by a static quark-antiquark pair. 
Simulations were done for unquenched SU(2) color with two flavors of staggered 
light quarks. An improved gluon action was used, which allows simulations to be 
done on coarse lattices, providing an extremely efficient means to access 
the large quark separations and long propagation times at which string breaking 
occurs. The static sources were generated using Wilson loop operators, hence no
light valence quarks are present in the resulting trial states. Results 
give unambiguous evidence of string breaking. First the static potential is 
shown to saturate at twice the heavy-light meson mass at large separations.
Then it is demonstrated that the local color-electric field strength in 
the region between the heavy quarks tends towards vacuum values at large 
separations, the first time that this most graphic effect of quark vacuum 
polarization on the confining flux-tube has been realized in lattice QCD. 
Implications of these results for unquenched 
simulations of four-dimensional QCD are drawn.
\end{abstract}

\maketitle

\section{Introduction}

The string model is widely used in phenomenologies of quark confinement 
and hadronization. A basic system of interest in the string model consists
of a heavy quark ($Q$) and antiquark ($\overline Q$) pair,
which are assumed to be bound by a narrow tube of color flux. This
model naturally gives rise to a linearly confining potential, if the 
tube has a cross-section that is approximately independent of the 
quark separation \cite{RichardHDT}.
In fact simulations of quenched lattice QCD long ago demonstrated the
formation of color-flux tubes having properties that are in accord 
with simple models.

In full QCD the confining $Q \overline Q$ string should not persist to arbitrarily 
large separations $R$, due to the effects of dynamical sea quarks. 
One expects that it becomes energetically favorable for a light quark ($q$) 
and antiquark ($\bar q$) to materialize from the Dirac sea at 
large $R$, and the light quarks should bind to the heavy quarks resulting in the 
formation of a pair of noninteracting color-neutral mesons. This phenomenon of 
hadronization, or string breaking, implies for instance that the potential 
$V(R)$ for a static $Q\overline Q$ pair should saturate at large separations,
\begin{equation}
   \lim_{R\to\infty} V(R) = 2M_{Q\bar{q}}, 
\label{PotentialSaturation}
\end{equation}
where $M_{Q\bar{q}}$ is the mass of the ground-state heavy-light meson. 

Observation of Eq.\ (\ref{PotentialSaturation}) in an unquenched 
lattice QCD simulation provides the simplest criterion to
demonstrate string breaking.
A much more detailed probe of this basic phenomenon of hadronic physics 
would come from measurements of the local color fields $f_{\mu\nu}(x;R)$ as
functions of position $x$ in the region between the static valence quarks, 
for various values of $R$. 
The effects of dynamical quarks should be to suppress the fields in the 
region between the static sources, with the fields saturating at 
vacuum values in the limit of large $R$
\begin{equation}
   \lim_{R\to\infty} f_{\mu\nu}(x;R) = 0 ,
\label{FieldSaturation}
\end{equation}
for $x$ outside a finite region surrounding each static valence quark,
corresponding to the size of the $Q\bar q$ meson. 

Despite considerable effort, however, it has proven to be very difficult to establish
Eq.\ (\ref{PotentialSaturation}), and especially Eq.\ (\ref{FieldSaturation}), 
in unquenched lattice QCD simulations  \cite{Reviews,Recent,FiniteT}.
The traditional computational strategy is to use Wilson loop operators 
in order to create trial states containing only static sources, with no valence 
light quarks.
In that case  Eqs.\ (\ref{PotentialSaturation}) and
(\ref{FieldSaturation}) provide unambiguous evidence of the effects of
light quarks that ``materialize'' from the sea (one must take care to note however 
that Euclidean lattice simulations only provide an analogy to the real-time 
process of hadronization).
The difficulty in making these measurements on the lattice is due in part
to the exponential suppression of the correlators, reflecting the large
energy of the state at large separations $R$, and the large propagation time 
$T$ which is necessary in order to isolate the ground state of the 
system \cite{HDT,Duncan,MixedOperators}.

A convincing demonstration of the saturation of the unquenched $V(R)$ 
has only recently been achieved, thanks in part to the realization
that improved actions can be used to do accurate simulations on lattices 
with coarse grids, where the computational effort can go into generating 
large ensembles of gauge-fields at the large length scales relevant to 
string breaking, rather than to
generating short-distance modes that are not of interest here \cite{HDT,Duncan}. 
Studies of the local color-fields are still more problematic than
the static potential; the simplest definition of $f_{\mu\nu}(x;R)$ 
comes from the correlator between the Wilson loop and the square of the 
local field strength, which requires a vacuum 
subtraction, leading to a very poor signal-to-noise ratio.
A demonstration that the fields in the region between the heavy quarks
become vacuum saturated has not been realized to date.

In this work we turn to Euclidean lattice QCD in three dimensions
(QCD$_3$), exploiting the computational advantage of this toy model 
(including a power law suppression of lattice discretization errors in this 
super-renormalizable theory) in order to convincingly demonstrate that 
dynamical quarks lead to saturation of the static potential
and the local color fields. This has been done using very little
computational power.

Lattice studies of confinement physics using QCD$_3$ have a long history; 
this theory exhibits most of the hallmarks of four-dimensional QCD,
including confinement, flux-tube formation, chiral symmetry breaking, 
a rich glueball spectrum, and a finite-temperature deconfining 
phase transition \cite{Teper}.

Some of the work presented here is a follow-up to an earlier study on
string breaking in unquenched QCD$_3$ by one of us \cite{HDT,KitThesis}. 
What is new includes a much more extensive study of the static potential 
on larger lattices and with higher statistics, enabling measurements of the 
potential out to substantially larger $R$. More important, in this work we 
provide the first demonstration in unquenched lattice QCD, as defined in any 
number of dimensions, of vacuum saturation of the local color fields.

Although these results are for a three-dimensional theory, they have
important implications for string breaking studies in QCD$_4$. 
In particular we find that string breaking is observed using trial states 
without light valence quarks, when they are propagated for Euclidean times 
$T \approx 1$~fm, longer than what has been attained in earlier 
unsuccessful studies, but not excessively so, this being the
natural scale associated with hadronic binding. 
[As discussed in Sect.\ \ref{sec:Simulations} we find a consistent
relation between the physical length scales in QCD$_3$ and in four-dimensional 
hadronic physics, using two very different quantities to identify 
the lattice spacing in QCD$_3$: the quenched string tension, 
and the lightest unquenched vector 
meson mass, analogous to the $\rho$.] 

Saturation in both the potential
(as previously observed in \cite{HDT}), and in the local color fields, 
is found to occur at separations $R \approx 1.5$~fm, which is similar to
estimates of the string breaking scale in QCD$_4$ \cite{QCD4Rb}.
The correlation functions are noisy, yet a sufficiently large ensemble of
configurations was generated with relatively little computational effort, 
even by the standards of the intrinsically cheaper three-dimensional theory, 
because we used improved actions to accurately simulate on coarse lattices, 
with spatial spacings $a_s \approx 0.2$~fm. 
The same reasoning should apply to the situation in 
unquenched QCD$_4$ \cite{Duncan}. 

The rest of this paper is organized as follows. In Sect.\ \ref{sec:Simulations} 
we describe the lattice actions and simulation parameters, and we also
discuss our scale setting procedure for making contact between QCD$_3$
and QCD$_4$. Results for the static potential are presented in 
Sect.\ \ref{sec:Potential} and results for the local field strength in 
Sect.\ \ref{sec:Fields}. A summary and some further discussion of the 
implications of this work for simulations of QCD$_4$ are found 
in Sect.\ \ref{sec:Summary}.

\section{\label{sec:Simulations}Simulations}

Unquenched simulations of SU(2) color in three dimensions were done using a 
tree-level $O(a^2)$-accurate improved gluon action, allowing for different
``temporal'' and ``spatial'' lattice spacings $a_t$ and $a_s$, respectively
\begin{equation}
S_{imp}= -\beta \sum_{x,\mu>\nu} \xi_{\mu\nu}
\left[ \frac{5}{3} P_{\mu\nu}
- \frac{1}{12} \left(R_{\mu\nu}+R_{\nu\mu}\right) \right],
\label{Sglue}
\end{equation}
where $P_{\mu\nu}$ is one-half the trace of the $1 \times 1$ plaquette and
$R_{\mu\nu}$ is one-half the trace of the $1 \times 2$ rectangle in the
$\mu$ $\times$ $\nu$ plane. The bare lattice anisotropy is input through
$\xi_{ij}= \xi_0 \equiv (a_t/a_s)_{\rm bare}$ [for $i,j=1,2$], and 
$\xi_{k3}=\xi_{3k}=1/\xi_0$. 
The three-dimensional theory is super-renormalizable, with a bare coupling $g_0^2$ 
having dimensions of mass, and which enters Eq.\ (\ref{Sglue}) through the 
dimensionless combination $\beta=4/(g_0^2 a_s)$.

We use the unimproved Kogut-Susskind staggered-quark action, which in three 
dimensions describes two flavors of four-component spinors \cite{Sks3d}
\begin{widetext}
\begin{equation}
S_{\rm KS} = 2am_0 \sum_x \bar\chi(x) \chi(x) 
+ \sum_{x,\mu} \zeta_\mu \eta_\mu(x) \bar\chi(x)
\left[ U_\mu(x) \chi(x+\hat\mu)
- U_\mu^\dag(x-\hat\mu) \chi(x-\hat\mu) \right],
\label{SKS}
\end{equation}
\end{widetext}
where $\eta_{\mu}(x)=(-1)^{x_{1}+\ldots+x_{\mu-1}}$ is the usual staggered
phase, and where $\zeta_{1,2}=1$, $\zeta_{3}=1/\xi_0$.

\begin{table}[htb]
\centering
\begin{tabular}{|c||c|c|}
\hline
Simulation Parameters   & Ref. \cite{HDT}   & This work         \\ \hline
Lattice Volume          & $16^{2}\times 10$ & $22^{2}\times 28$ \\ \hline
$\beta$                 & 3.0               & 3.0               \\ \hline
$(a_t/a_s)_{\rm bare}$  & isotropic         & 1/2               \\ \hline
$m_0/g_0^2$             & 0.075             & 0.10              \\ \hline
$N_{\rm meas}$          & 6,000             & 30,000            \\ \hline
\end{tabular}
\caption{Comparison of simulation parameters used in an earlier 
study by one of us \cite{HDT}, and in the present work. 
The quenched and unquenched simulations in this work 
were done at the same couplings. $N_{\rm meas}$ is
the number of measurements in each unquenched study.}
\label{table:SimParams}
\end{table}

Simulations were done on a $22^{2}\times 28$ lattice at $\beta=3$, with
a bare anisotropy $\xi_0 = 1/2$, and a bare quark mass $m_0/g_0^2 = 0.1$. 
We find that this corresponds to a rather large pion mass, with 
$m_\pi / m_\rho \approx 0.61$. Hence the light quark used here
is actually rather heavy, comparable to the strange quark;
nevertheless this quark mass is apparently light enough to resolve 
string breaking at accessible separations $R$.
These parameters are very similar to those that were used in Ref. \cite{HDT}, 
although the lattice volume is substantially larger here. For comparison the 
simulation parameters in the present study are compared with those 
in Ref.\ \cite{HDT} in Table \ref{table:SimParams}. 

The configurations were generated using the hybrid molecular dynamics
algorithm (the $\Phi$-algorithm) \cite{HMDpaper} with time step
$\Delta t=0.02$ and 50 molecular dynamics steps taken for each
trajectory. Direct calculation of autocorrelation times for the largest 
Wilson loops showed $\tau \lesssim 0.2$ and, based on binning studies of
autocorrelations, we elected to skip 10 trajectories between measurements. 
Our final data set consists of 30,000 measurements, corresponding to a
total of 300,000 trajectories.

The bare parameters $g_0$, $m_0$, and $\xi_0$ are finite in the continuum 
limit of this super-renormalizable theory. At finite lattice spacings these
parameters must absorb finite renormalizations in order to keep physical
quantities fixed. However in Refs.\ \cite{HDT,HDTLat98} it was found that 
little renormalization is required when the improved gluon action is 
employed, in the range of couplings used here. 
For example, the unquenched static potential showed no discernable change 
from $\beta=3$ to $\beta=2$, corresponding to a 50\% increase in 
lattice spacing \cite{HDT}; meson masses were also found to have 
small discretization errors \cite{HDTLat98}. Of relevance to
our comparison of the quenched and unquenched results is the fact
that the lattice spacings in the two theories are found to be very similar 
at a given bare coupling $\beta$ (a quantitative comparison is made below). 
Moreover one finds similar meson masses in the two theories,
at a given bare quark mass; for example we find $m_\pi / m_\rho \approx 0.75$
in the quenched theory, compared to the value $\approx 0.61$ unquenched.

We likewise find very little renormalization of the bare lattice anisotropy.
The renormalized anisotropy was determined by comparing measurements of 
the static potential where the direction of Euclidean time was oriented along
the $a_t$ lattice axis and then along an $a_s$ axis. The results of the
quenched simulations give a renormalized value $a_t / a_s = 0.478(7)$,
while in the unquenched theory we find $a_t / a_s = 0.455(9)$ 
(much larger renormalizations are found if the unimproved Wilson 
gluon action is used \cite{HDT}). 

One can make rough contact between the length scales relevant to 
string breaking in QCD$_3$, and the corresponding scales 
in four-dimensional QCD, by identifying a confining scale in the two theories 
\cite{HDT}.  For instance, one can identify the quenched string tension in QCD$_4$, 
$\sqrt\sigma\approx0.44$~GeV, with the result in quenched three-dimensional 
SU(2) color \cite{Teper}, $\sqrt{\sigma}/g_0^2=0.3353(18)$.
In effect this allows one to express the dimensionful coupling $1/g_0^2$, or 
equivalently the lattice spacing $a_s$ at a given $\beta$, in the three-dimensional 
theory in ``fermi.'' For instance at $\beta=3$ one finds $a_s \approx 0.20$~fm.
The consistency of this procedure can be established by identifying other
physical quantities in QCD$_3$ and QCD$_4$. We measured the mass $m_V$
of the lightest vector meson, with the result $a_t m_V = 0.425(8)$ 
in the unquenched case, and identifying this with the physical $\rho$ meson 
mass one finds $a_s \approx 0.24$~fm.
In the quenched theory at the same couplings the vector meson mass is 
$a_t m_V = 0.380(2)$, which would identify the spacing as $a_s \approx 0.20$~fm. 
[This also supports the contention that the quenched and unquenched QCD$_3$ 
theories can be directly compared at the same bare couplings, though a small
mismatch in the couplings will not affect the trends that we will use to 
establish Eqs.\ (\ref{PotentialSaturation}) and (\ref{FieldSaturation}).]

It is important to note at this point that one can, and we will, establish 
the string breaking criteria completely within the QCD$_3$ theory, independent of 
this identification of scales with QCD$_4$. This connection is only used 
to the extent that it gives one some physical intuition for the length scales
that are being probed in QCD$_3$, and since one may also thereby make some comparison 
between the parameters of the simulations that are done here, which successfully 
resolve string breaking, and those that have been used in various studies 
in QCD$_4$. It is to these ends that we sometimes label our simulation 
parameters according their values in ``fermi.'' 
We return to the comparison of our simulation results with QCD$_4$ in 
Sect.\ \ref{sec:Summary}.

\section{\label{sec:Potential}Results: Static Potential}

We extracted the static potential from expectation values $\langle W(R,T) \rangle$
of $R\times T$ Wilson loops.
Standard APE-type smearing was used for the spatial links in the Wilson loops, 
in order to enhance the overlap with the ground state. Measurements were done both 
for on-axis and many off-axis separations (i.e.\ using link paths joining lattice 
sites displaced along both spatial directions).

\begin{figure}[htb]
\centering
\includegraphics[angle=90,width=\columnwidth]{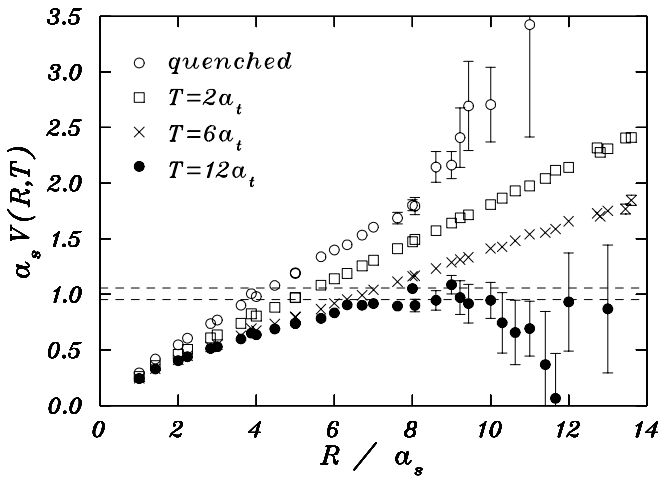}
\caption{Time-dependent unquenched effective potential $V(R,T)$ 
versus quark separation $R$, at several propagation times $T$.
The quenched potential is also shown. The dashed lines give 
the unquenched simulation measurement of $2 M_{Q\bar q}$.}
\label{fig:Potential}
\end{figure}

In order to demonstrate the onset of string breaking it is essential to 
carefully study the time-dependent effective potential $V(R,T)$ as a function of
the propagation time $T$, where 
\begin{equation}
V(R,T) \equiv  -\ln { \langle W(R,T) \rangle \over \langle W(R,T-1) \rangle } , 
\end{equation}
with the true ground state energy $V(R)$ given as usual by 
$V(R) = \lim_{T\to\infty} V(R,T)$. Results for the unquenched 
effective potential are presented in Fig.\ \ref{fig:Potential} 
for several values of $T$, and are compared with the quenched potential 
(the latter is evaluated at $T=12 a_t$, which is generally large enough to 
isolate the ground state in that case, except at the largest $R$'s in 
the figure). The potential is also compared with the unquenched 
simulation determination of twice the heavy-light meson mass 
(using an appropriate staggered interpolating field \cite{HeavyLight}).

These results clearly demonstrate string breaking as defined by 
Eq.\ (\ref{PotentialSaturation}). An asymptote in the unquenched potential at 
large separations is progressively revealed as $T$ increases; at the longest
propagation time shown ($T=12 a_t \approx 1.2$~fm), the potential has clearly 
saturated at the expected limit of $2 M_{Q\bar q}$ over a very wide range of radii. 
From these results one can make a rough determination of the separation
$R_{\rm sb}$ for the onset of string breaking, corresponding to 
the point at which the potential reaches the correct asymptote. We find 
$R_{\rm sb} \approx \mbox{(6--8)}a_s$. Using the scale setting procedure
described in Sect.\ \ref{sec:Simulations} to convert the lattice spacing
to ``fermi,'' one obtains $R_{\rm sb} \approx \mbox{(1.2--1.6)}$~fm,
which is very similar to estimates of the string breaking distance 
in QCD$_4$ \cite{QCD4Rb}. 

\begin{figure}[htb]
\centering
\includegraphics[angle=90,width=\columnwidth]{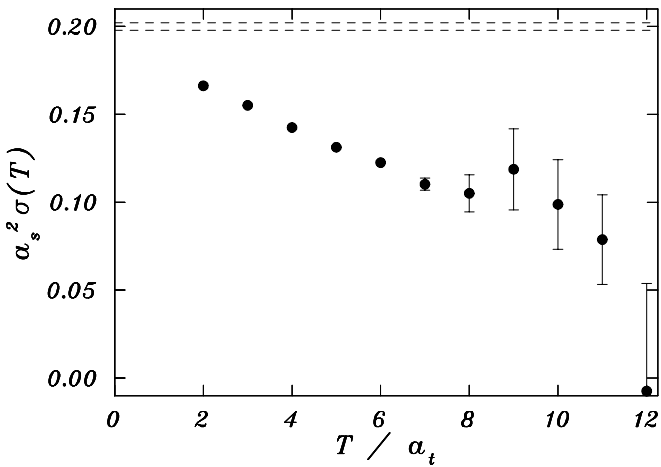}
\caption{Time-dependent unquenched ``effective'' string tension 
$a_{s}^{2}\sigma(T)$, defined according to Eq.\ (\ref{sigmaT}). 
The dashed lines show the quenched string tension.}
\label{fig:StringTension}
\end{figure}

To further analyze the approach to an asymptote in the effective potential as 
$T$ is increased, we define an ``effective'' string tension $\sigma(T)$,
computed from $V(R,T)$ over a small interval in separation
$R$ that is beyond the string breaking distance $R_{\rm sb}$; we take
\begin{equation}
   V(R,T) = \sigma(T) R + b, \quad R/a_s \in[8,10] .
\label{sigmaT}
\end{equation}
Results for $\sigma(T)$ so defined were obtained from fits to the simulation 
measurements, and are shown in Fig.\ \ref{fig:StringTension}. 
The unquenched string tension clearly tends towards zero with increasing $T$, 
and this trend is manifest even at the smallest propagation times,
where the measurements are most accurate. An extrapolation by-eye of the 
trend in $\sigma(T)$ from the smallest times would suggest that the string
tension vanishes at a propagation time that is only slightly longer than
what is revealed by direct inspection of an asymptote in $V(R,T$) at large $T$, 
which clearly sets in by $T=12a_t$, as seen in Fig.\ \ref{fig:Potential}.

\section{\label{sec:Fields}Results: Local color fields}

The simplest observable for measuring the local color fields comes
from the correlator of a Wilson loop $W(R,T)$ with the plaquette
$P_{\mu\nu}$
\begin{equation}
   f_{\mu\nu}(\vec x) = -{\beta \over a^3} \left[
     {\langle W(R,T) P_{\mu\nu}(\vec x) \rangle \over \langle W(R,T) \rangle}
    - \langle P_{\mu\nu} \rangle \right] ,
\end{equation}
where we suppress the dependence of $f_{\mu\nu}$ on $R$ and $T$. We measure
$\vec x$ relative to the center of the Wilson loop.
In the continuum limit (and at sufficiently large $T$) the above correlator 
corresponds to the expectation value of the square of the Euclidean field 
strength in the presence of the static $Q \overline Q$ pair, after vacuum
subtraction
\begin{equation}
   \lim_{a\to 0 \atop  T\to\infty} f_{\mu\nu}(\vec x)
   = {\left\langle \mbox{Tr} \left(F_{\mu\nu}(\vec x)\right)^2 \right \rangle}_{Q\overline Q} 
   - {\left\langle \mbox{Tr} \left(F_{\mu\nu}\right)^2 \right \rangle}_{\rm vac} ,
   \end{equation}
where $F_{\mu\nu}$ denotes the usual continuum field strength, and
$\langle \ldots \rangle_{Q\overline Q}$ denotes the expectation value in
the presence of the static sources, and $\langle \ldots \rangle_{\rm vac}$ 
is the vacuum expectation value.

The most important component of the field correlator corresponds to the
color-electric field with vector component along the direction parallel
to the axis joining the quark and antiquark (which, for instance,
makes the largest contribution to the potential energy).
We denote this correlator by $E^\parallel(\vec x)$, given by
\begin{equation}
   E^\parallel(\vec x) = -f_{1t}(\vec x) ,
\end{equation}
where the minus sign accounts for the Euclidean contribution to the energy
density, and where here the ``1'' direction denotes the $Q\overline Q$ axis and $t$
denotes the Euclidean time axis corresponding to the small lattice spacing $a_t$.

\begin{figure}[tb]
\centering
\subfigure{\includegraphics[angle=90,width=\columnwidth]{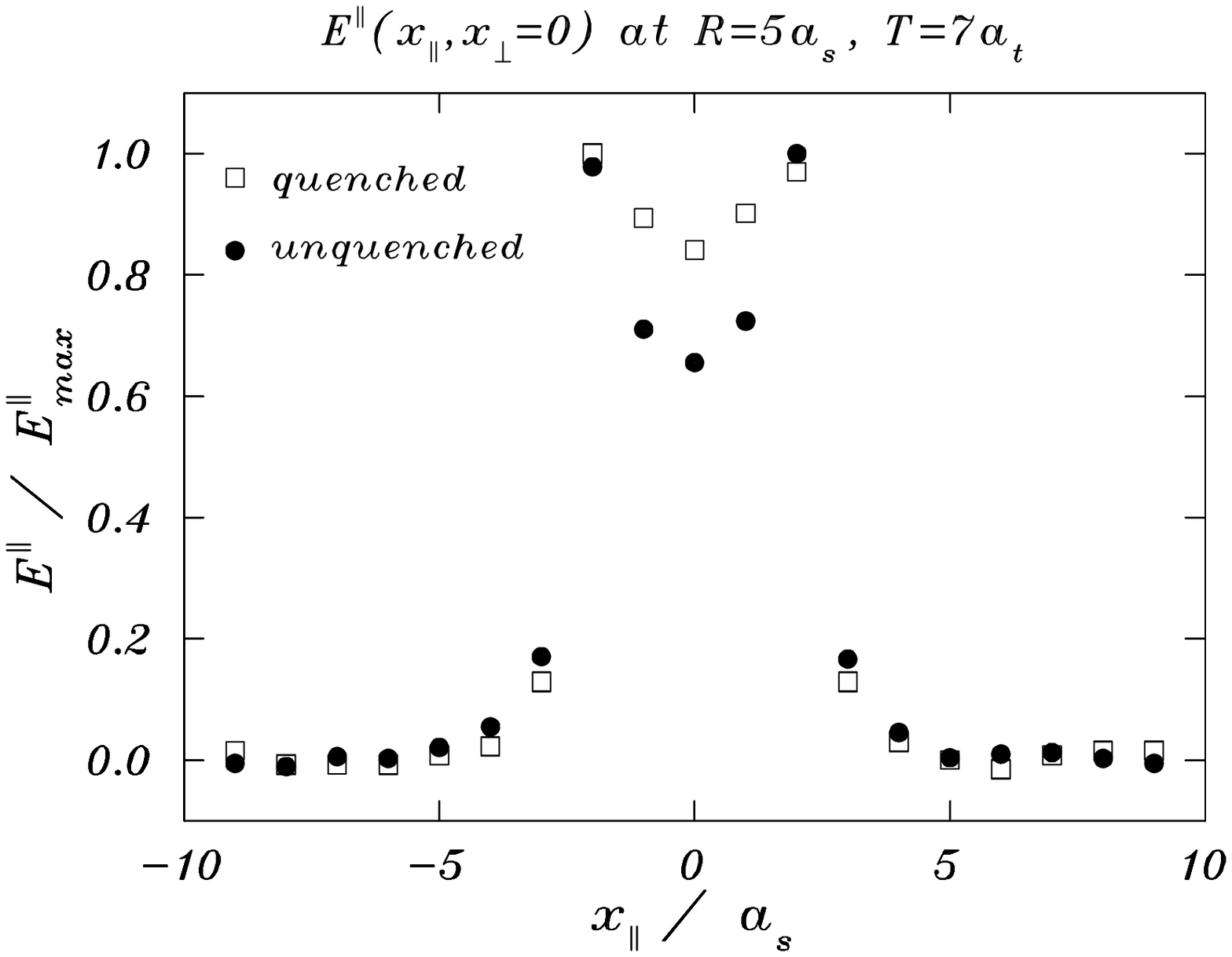}}
\subfigure{\includegraphics[angle=90,width=\columnwidth]{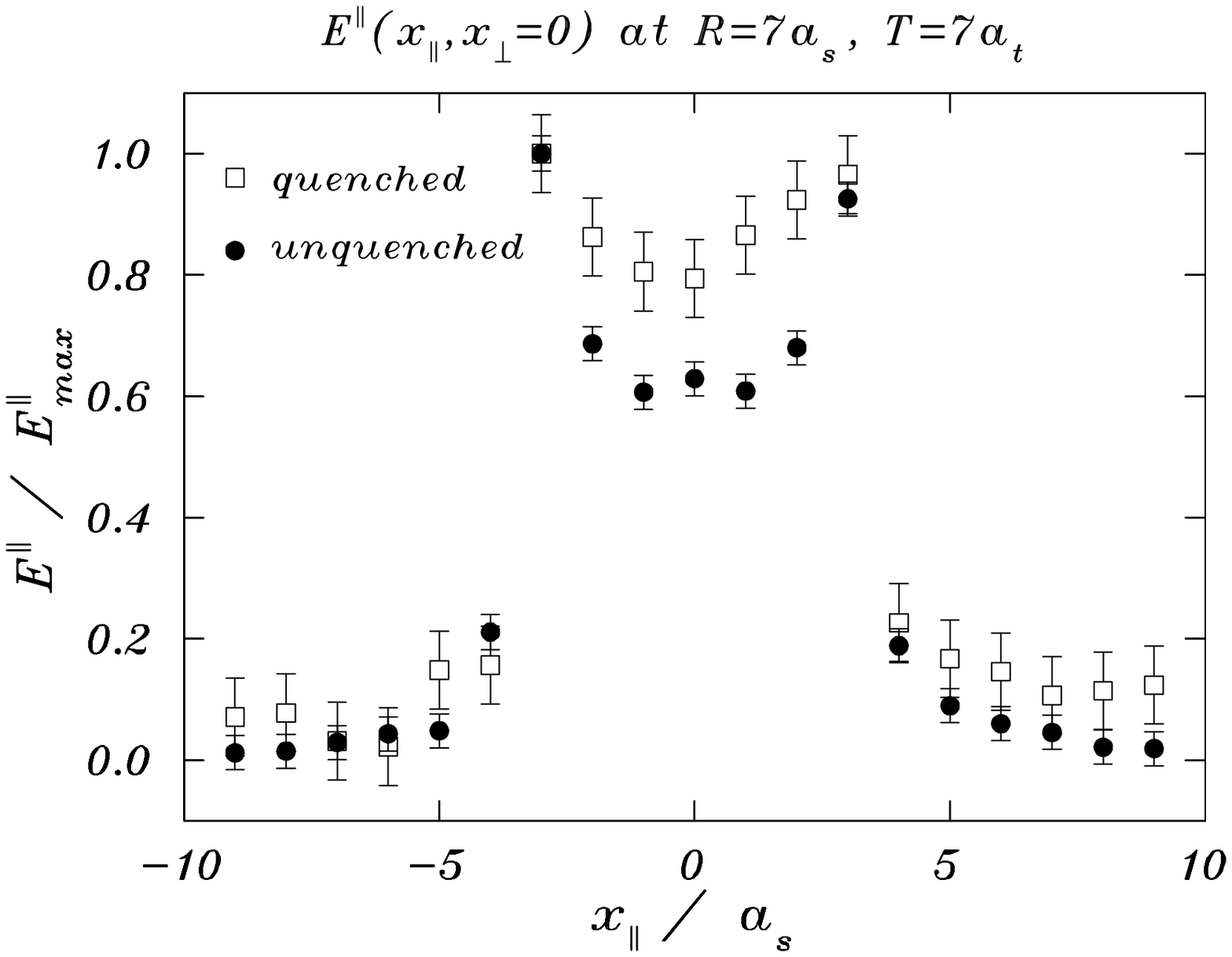}}
\caption{Color-electric field $E^\parallel(x_\parallel,x_\perp=0)$ as a function
of position $x_\parallel$ along the axis joining the quark and antiquark, 
for two values of the separation, at a fixed $T$. 
$E^\parallel_{\rm max}$ is the maximum field strength, in the given theory, 
along this line (i.e.\ the field at the location of either quark).}
\label{fig:xPara}
\end{figure}

\begin{figure}[tb]
\centering
\subfigure{\includegraphics[angle=90,width=\columnwidth]{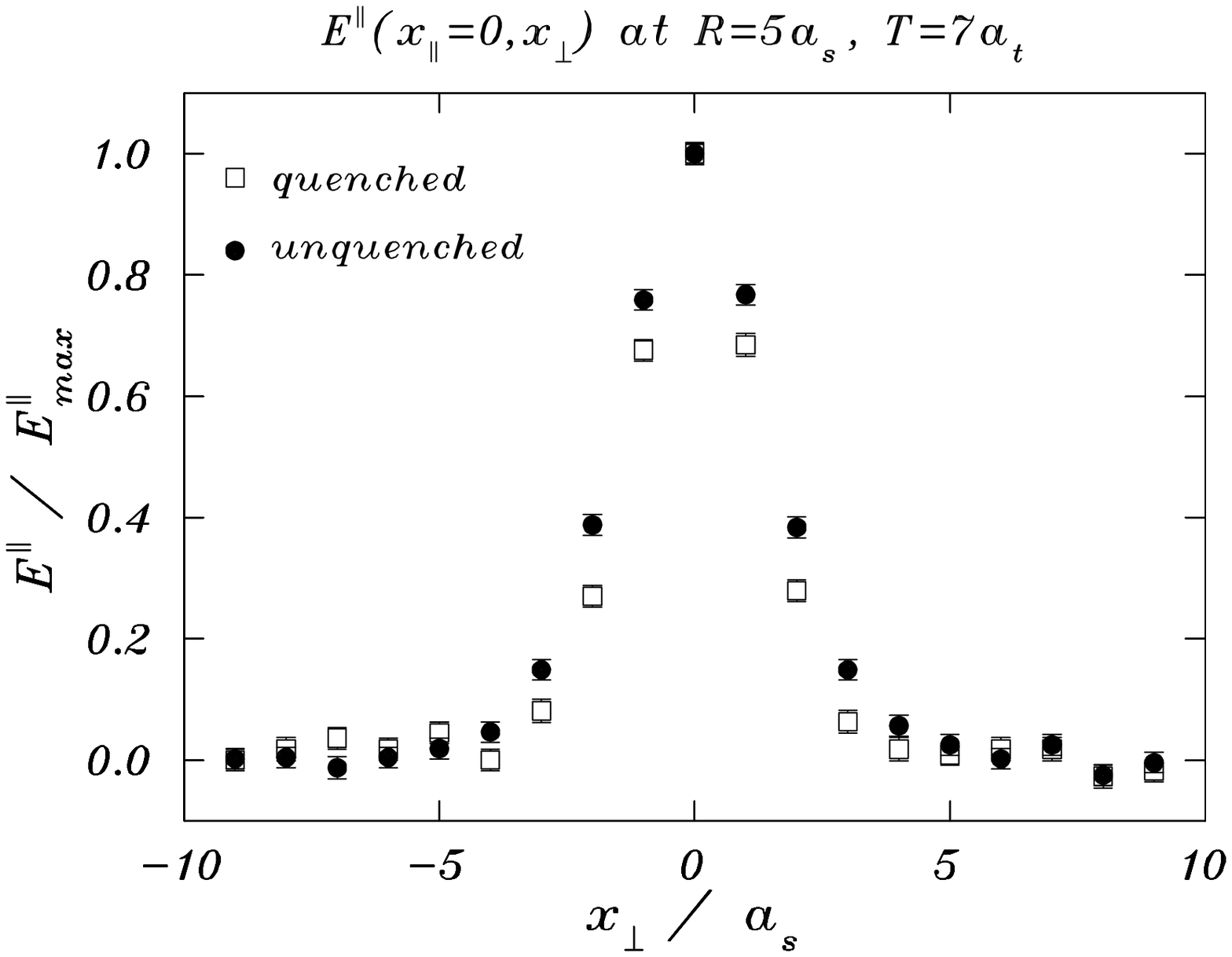}}
\subfigure{\includegraphics[angle=90,width=\columnwidth]{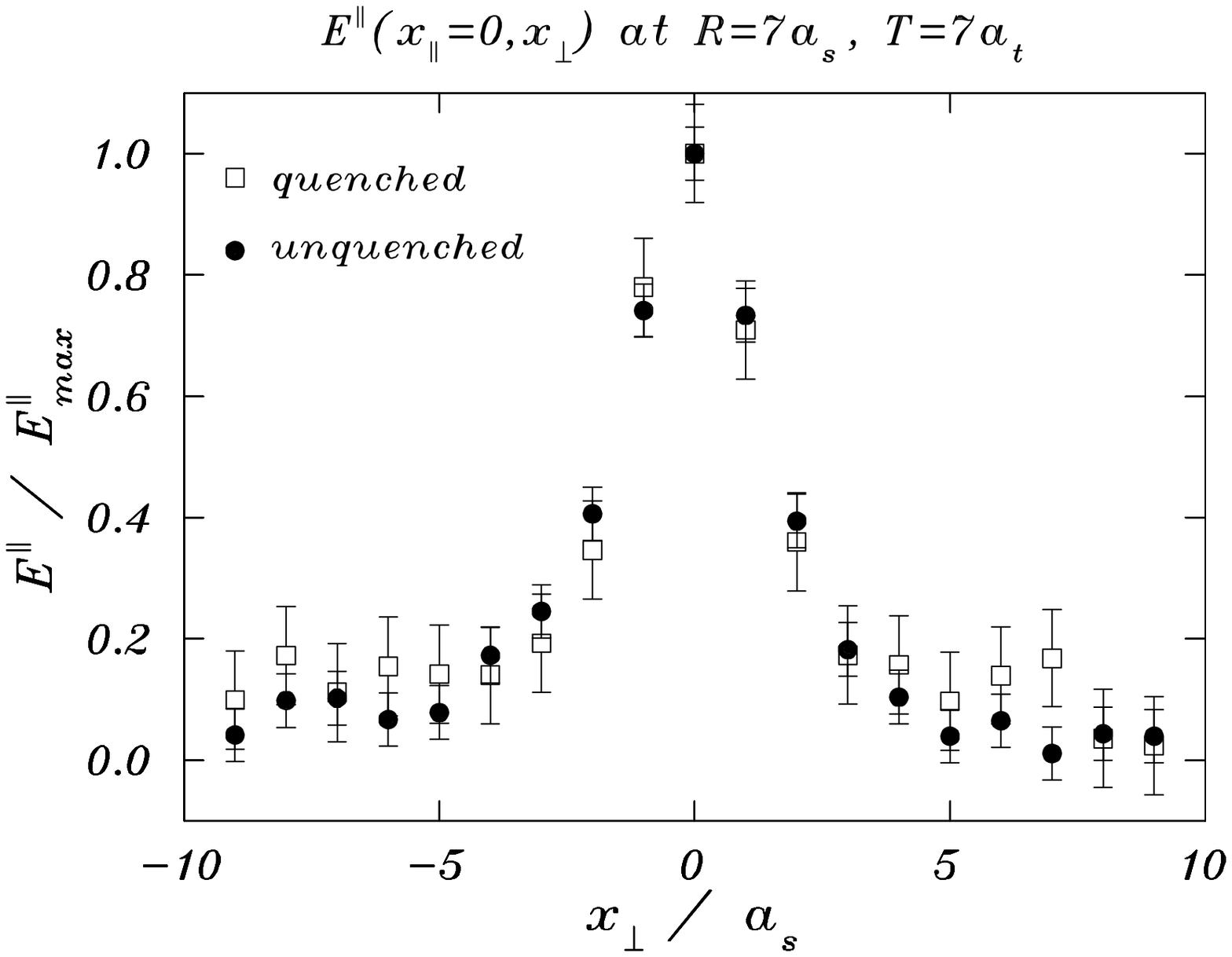}}
\caption{Color-electric field $E^\parallel(x_\parallel=0,x_\perp)$ as a function of
position $x_\perp$ along a line perpendicular to the axis joining the quark and antiquark, 
for two values of the separation, at a fixed $T$. 
$E^\parallel_{\rm max}$ is the maximum field strength, in the given theory, 
along this line (which in this case corresponds to the field at the midpoint 
of the $Q\overline Q$ axis).}
\label{fig:xPerp}
\end{figure}

We present one-dimensional profiles of the field strength, along two different
lines through the flux-tube, in Figs.\ \ref{fig:xPara} and \ref{fig:xPerp}. 
We write $\vec x = (x_\parallel, x_\perp)$, where 
$x_\parallel$ is the position along the central axis joining the quarks, 
and $x_\perp$ is the position perpendicular to that line, both measured 
from the center of the axis.
We plot the profile along the central axis joining the quarks in
Fig.\ \ref{fig:xPara}, and along a line perpendicular to the central axis in
Fig.\ \ref{fig:xPerp}. The quenched and unquenched correlators are shown in
each case and, in order to emphasize the shape of the profiles in the two
theories, each correlator is normalized to its maximum value along 
the particular line. 

The existence of a flux-tube is demonstrated in part by a ``plateau'' 
in the field strength in a region between the sources 
(which is expected to increase in width as $R$ increases),
where the field is approximately independent of $x_\parallel$, as
seen in Fig.\ \ref{fig:xPara}. The relative strength of the field in this region 
is suppressed in the unquenched theory. 
[We note that $E^\parallel$ in the plaquette nearest to
a static source, at $\vert x_\parallel \vert = {\rm int}(R/2)$, $x_\perp=0$, 
is generally about half as large in the unquenched theory as in the quenched case,
at large $R$. To the extent that the couplings in the two theories are 
well matched, this already illustrates the suppression of the fields due
to the effects of unquenching, since the correlator is an average of
the squared-field over a unit cell.]

Notice as well the approximate symmetry of the flux-tube profiles about the center
of the $Q\overline Q$ axis. The transverse widths of the flux tubes can be
inferred from Fig.\ \ref{fig:xPerp}, and are indistinguishable
in the unquenched and quenched cases (the width is also roughly independent 
of separation, which is consistent with simple models \cite{RichardHDT}). 

\begin{figure}[htb]
\centering
\includegraphics[angle=90,width=\columnwidth]{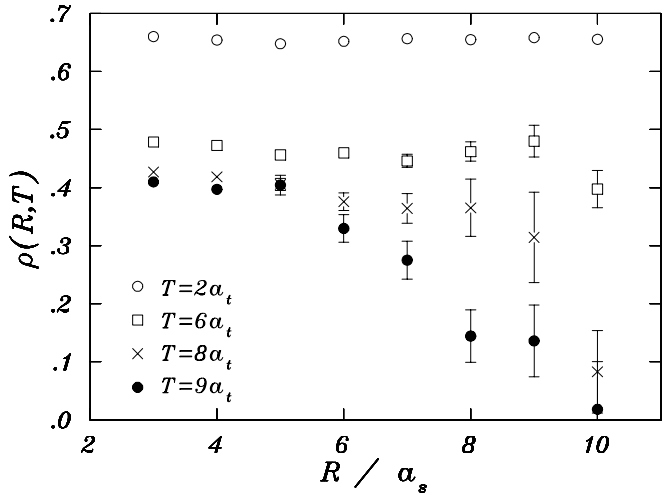}
\caption{Ratio of the field strengths, unquenched to quenched, at the 
midpoint of the $Q\overline Q$ axis, as a function of separation $R$, 
for various propagation times $T$.}
\label{fig:FluxRatiovsR}
\end{figure}

\begin{figure}[htb]
\centering
\includegraphics[angle=90,width=\columnwidth]{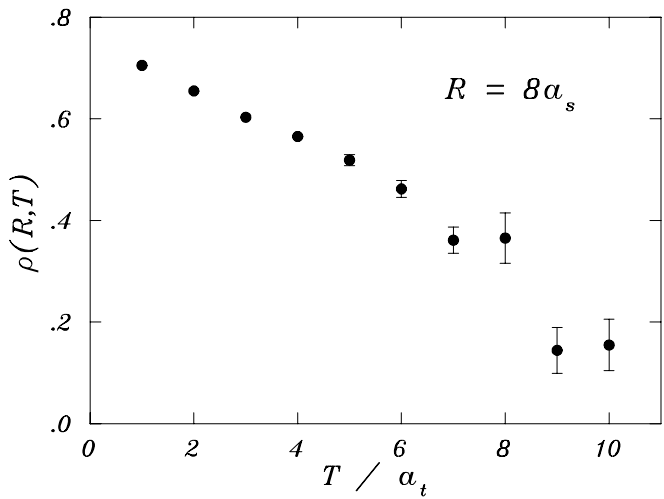}
\caption{Ratio of the field strengths, unquenched to quenched, at the
midpoint of the $Q\overline Q$ axis, as a function of propagation time $T$, 
for fixed $R=8a_s$.}
\label{fig:FluxRatiovsT}
\end{figure}

The results in Figs.\ \ref{fig:xPara} and \ref{fig:xPerp} are shown for fixed
$T=7a_t\approx 0.7$~fm. As indicated by the analysis of the potential in the 
previous section, one can easily be misled as to the extent of string breaking
unless one attains propagation times of at least 1~fm. This is particularly
difficult in the case of the field correlators. Alternatively one can
measure at several ``intermediate'' values of $T$, and study the
systematics of increasing propagation times. To illustrate this we compare
the fields at the midpoint of the $Q\overline Q$ axis in the two theories
 \begin{equation}
   \rho(R,T) \equiv 
   \left[ E^\parallel(\vec x = 0) \right]_{\rm unquenched}
   \Bigl/ \left[ E^\parallel(\vec x = 0) \right]_{\rm quenched} ,
\end{equation}
where we explicitly indicate the dependence of the ratio on $R$ and $T$.
The results are shown in Figs.\ \ref{fig:FluxRatiovsR} and \ref{fig:FluxRatiovsT},
which provide analogues for the fields to the effective potential plots 
in Figs.\ \ref{fig:Potential} and \ref{fig:StringTension}.

The unquenched fields are suppressed at all $R$ and $T$, however at modest
propagation times the unquenched field does not drop with
increasing $R$; taken at face value, this might suggest a failure to
observe string breaking, perhaps due to some pathology in the trial
states generated by Wilson loops, as has been suggested by other
authors (see e.g.\ Refs. \cite{CPPACS,MixedOperators}). In 
actuality this is simply a reflection of the fact that the ground state is not 
resolved unless one achieves propagation times $\approx 1$~fm, which is 
characteristic of hadronic binding. 

The systematics are particularly well illustrated in 
Fig.\ \ref{fig:FluxRatiovsT}, which shows the time-dependence of the 
field strength for a fixed separation $R=8 a_s$, which was chosen because it
is just beyond the string breaking distance $R_{\rm sb}$. The trend 
is once again manifest even at small propagation times, where the 
signal is very clean, and an extrapolation by-eye indicates that the 
field becomes vacuum saturated for $T \agt 12 a_t$, which is consistent 
with our analysis of the effective potential in the previous section.

\section{\label{sec:Summary}Summary and Outlook}

We were able to observe string breaking by dynamical quarks in
three-dimensional QCD by working on a coarse lattice, using the staggered 
quark action along with an improved gluon
action. String breaking was observed at large quark-antiquark separations,
both as saturation of the static potential at twice the heavy-light meson mass, 
according to Eq.\ (\ref{PotentialSaturation}), and as saturation of the 
local field strength at vacuum values, according to Eq.\ (\ref{FieldSaturation}). 
Measurements were done using only trial states generated by Wilson loop operators, 
in which no valence light quarks are present. Hence saturation of the 
unquenched potential and fields are highly nontrivial results, 
due entirely to the effects of dynamical sea quarks. 
The local field strength is particularly difficult to compute at the propagation
times and separations at which string breaking occurs. Nonetheless computations
were done over a range of propagation times that was sufficient to clearly 
establish the trend for the field strength to approach vacuum values in
the region between the static sources.

Although this work was done in QCD$_3$, the results provide clear implications
for large scale simulations of full QCD in four dimensions \cite{HDT,Duncan}.
Of particular importance is the need to achieve sufficiently long propagation times 
for the correlation functions; one can easily be misled as to the extent of 
string breaking unless one attains propagation times of at least 1~fm,
the scale characteristic of hadronic binding. 
It is also sensible to work on relatively coarse lattice, using
improved actions to achieve accurate results, since this alleviates
the computational burden associated with generating
short-distance modes that are not relevant to string breaking.  

\begin{table}[htb]
\centering
\begin{tabular}{|c||c|c||c|}
\hline
     & QCD$_4$ Ref.\ \cite{CPPACS} & QCD$_4$ Ref.\ \cite{Duncan} 
     & QCD$_3$ (this work) \\ \hline
Eq.\ (\ref{PotentialSaturation})?
     &  No                     & Yes
     & Yes                                       \\ \hline
$a_{s}$        
     & 0.15~fm                 & 0.40~fm
     & 0.20~fm                                   \\ \hline
$m_\pi/m_\rho$ 
     & 0.59                    & 0.25
     & 0.61                                      \\ \hline
$T_{\rm max}$    
     & 0.4--0.8~fm             & 1.2~fm
     & 1.2~fm                                    \\ \hline
$R_{\rm max}$    
     & 1.8~fm                  & 1.2~fm
     & 2.8~fm                                    \\ \hline
$N_{\rm meas}$
     & 492                     & 1,000
     & 30,000                                    \\ \hline
Volume      
     & $16^3 \times 32$        & $6^4$    
     & $22^2 \times 28$                          \\ \hline
\end{tabular}
\caption{Comparison of some recent unquenched lattice simulations in 
QCD$_4$ and in the present QCD$_3$ analysis, distinguished in part by whether 
the particular study resolved string breaking in the static potential according 
to Eq.\ (\ref{PotentialSaturation}). The lattice spacings, meson masses, 
maximum quark separations $R_{\rm max}$ and correlation function propagation times 
$T_{\rm max}$, number of measurements $N_{\rm meas}$, and lattice
volumes are compared.}
\label{table:QCD4and3}
\end{table}

In this connection we compare the parameters of two simulations 
of unquenched QCD$_4$ (which used trial states generated only by
Wilson loops) and of the present study of QCD$_3$, 
in Table \ref{table:QCD4and3}. These studies are distinguished in part
by whether the string breaking condition Eq.\ (\ref{PotentialSaturation})
was resolved. One must exercise caution when comparing the
two simulations in QCD$_4$, which were done with different actions 
and simulation algorithms.
One must likewise exercise caution in the comparison between the 
QCD$_4$ and QCD$_3$ simulations (taking note of the discussion at the end of 
Sect.\ \ref{sec:Simulations} concerning the scale setting procedure
that is used for this purpose).
Nonetheless this comparison is instructive, as it highlights the advantages
to be gained by working on coarser lattices. In unquenched QCD$_4$ the 
computational cost scales with lattice spacing roughly as $a^{-7}$, 
hence a small increase in the coarseness of the lattice can significantly reduce
the computational cost, allowing one to generate much larger ensembles and/or
to work at much smaller quark masses. 
Indeed the QCD$_4$ study in Ref.\ \cite{Duncan}, which did resolve string
breaking in the static potential, only required the equivalent of few 
PC-years of run time (comparable to the cost of the present study).

In this work we balanced the desire to use the coarsest lattice 
possible, in order to minimize the computational cost, against the need to 
have a spacing that is fine enough to resolve the spatial distribution 
of the color fields. We also systematically studied the trends in correlation 
functions with increasing propagation times, in order to leverage the
cleaner data at smaller $T$. It is reasonable to anticipate that
string breaking in the local color fields, Eq.\ (\ref{FieldSaturation}),
may be accessible to simulations in QCD$_4$, following the approach that
proved successful here in QCD$_3$.

\acknowledgments
We thank Richard Woloshyn and Peter Lepage for helpful discussions.
The work was supported in part by the Natural Sciences and Engineering
Research Council of Canada.


\end{document}